\documentclass[showpacs,preprintnumbers,amsmath,amssymb,superscriptaddress,twocolumn]{revtex4}
\usepackage{graphicx}
\usepackage{dcolumn}
\usepackage{bm}

\def\sgn{\mathop{\rm sgn}}

\begin{document}

\title{Gradient Echo Quantum Memory for Light using Two-level Atoms }

\author{G.~H\'etet}
\affiliation{ARC COE for Quantum-Atom Optics, Australian National
  University, Canberra, ACT 0200, Australia}

\author{J.~J.~Longdell}
\affiliation{Laser Physics Centre, RSPhysSE, Australian National
  University, Canberra, ACT 0200, Australia}
\affiliation{Department of Physics, University of Otago, Dunedin, New
Zealand}
\author{A. L. Alexander}
\affiliation{Laser Physics Centre, RSPhysSE, Australian National
  University, Canberra, ACT 0200, Australia}

\author{P. K. Lam}
\affiliation{ARC COE for Quantum-Atom Optics, Australian National
  University, Canberra, ACT 0200, Australia}

\author{M. J. Sellars}
\email[Email: ]{matthew.sellars@anu.edu.au}
\affiliation{Laser Physics Centre, RSPhysSE, Australian National
  University, Canberra, ACT 0200, Australia}

\begin{abstract}
We propose a quantum memory for light that is analogous to the NMR gradient echo.  Our proposal is ideally perfectly efficient and provides simplifications to current 3-level quantum memory schemes based on controlled inhomogeneous broadening.  Our scheme does not require auxiliary light fields.  Instead the input optical pulse interacts only with two-level atoms that have linearly increasing Stark shifts.  By simply reversing the sign of the atomic Stark shifts, the pulse is retrieved in the 
forward direction.  We present analytical, numerical and experimental results of this scheme.  We report experimental efficiencies of up to 15\% and suggest simple realizable improvements to significantly increase the efficiency.
\end{abstract}
\maketitle


Some of the most significant advances in quantum information
processing have been made using quantum optics. To
extend these techniques, it is necessary to have devices such as single photon
sources, quantum memories and quantum repeaters, where quantum
information is exchanged in a controlled fashion between light fields
and material systems. 

A quantum memory for light is a device that can efficiently delay or store the quantum states of light fields.  This is usually achieved via some form of imprinting onto an atomic system.  The quantum states stored by the device must also be faithfully retrievable on demand where the total efficiency of the processes must exceed the classical benchmark so that quantum information can be retained \cite{hamm05}.  It has been proposed that the requisite control and strong coupling can both be achieved using an ensemble approach, where the light field interacts with a large number of identical atoms \cite{Koz}.
In the discrete variable regime, single photon states have been stored and retrieved from magneto-optical traps using electromagnetically induced transparency \cite{chan05,eisa05}.  Whilst in the continuous variable regime, quantum states of light have been mapped onto atoms using the resonant interaction of light with spin polarised cesium vapors \cite{juls04}.  Nevertheless, further improvements on the efficiencies are still needed for realizing a reliable quantum memory.

\begin{figure}[!ht]
  \centering
\includegraphics[width=\columnwidth]{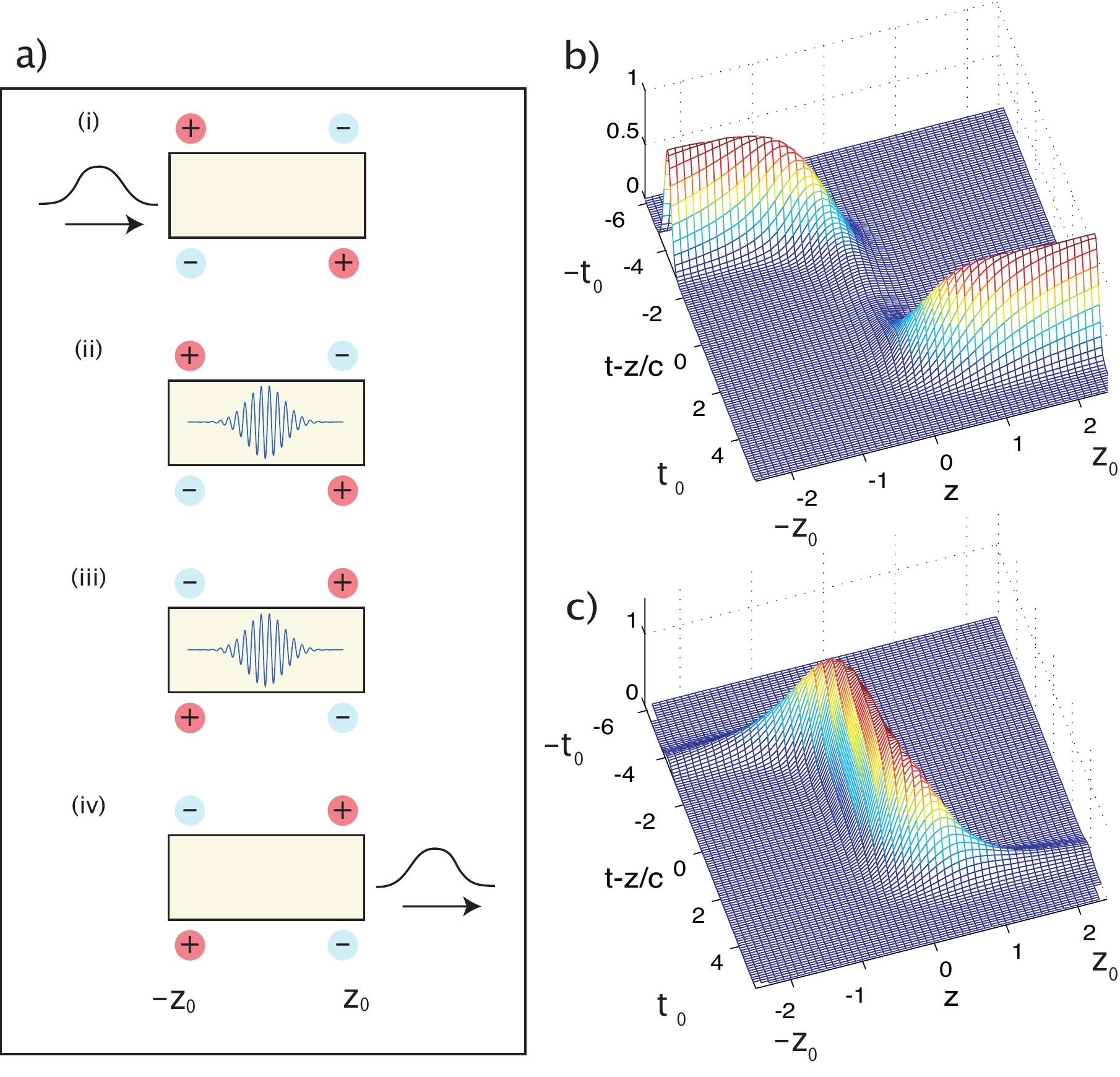}
  \caption{(Color online)  a) Schematic of a gradient echo quantum memory.  (i) The {\it input} light field enters the sample at $z=-z_0$ and $t=-t_0$; (ii) the optical information of the light field is {\it imprinted} onto the inhomogenously broadened two level atomic ensemble, (iii) to {\it recall} the light field, the polarity of the quadrupole electric field is flipped at $t=0$, (iv) finally an {\it output} field is {\it retrieved} at $z = z_0$ and $t = t_0$.  With the optical depth chosen to be $gN = 3.3 \eta$ and with no decoherence $\gamma = 0$, b) and c) show the space-time grid plots of the light field intensity and the atomic polarization, respectively.  The input pulse duration is $t_{\rm pulse} = t_0/4$ and the quadrupole induced broadening is $2/t_{\rm pulse}$.}
  \label{fig:mbsim}
\end{figure}

In 2001, Moiseev and Kr\"oll \cite{mois01} published a proposal for a
quantum memory for light based on modified photon echoes.  In contrast
to a normal photon echo, the rephasing came from controlled
reversible inhomogeneous broadening (CRIB) rather than from optical
pulses.  In the initial proposal the reversible inhomogeneous
broadening used opposite propagation directions in a Doppler broadened
medium.  Since then the idea has been generalised to other broadening
mechanisms \cite{krau06,nils05,mois03}.

Photon echoes using reversible inhomogeneous broadening \cite{alex06}, 
with the ability to store multiple pulses,  has been
demonstrated using Stark shifts in europium dopants \cite{alex07}.  Work towards
demonstrating such echoes in other systems has also been carried out
\cite{sang06,stau06,nils05}.  To date all the proposals for a quantum memory using CRIB operate via
a time reversal of the storage process.  By reversing the detunings of the
atoms, one can transform the equations of motion for light travelling in the
backward direction into a time reversed copy
travelling in the forward direction. A forward propagating input pulse is absorbed by the atomic ensemble.  Then the detunings
of the atoms are flipped and a phase matching operation is applied. The phase matching operation
consists of a pair of counter propagating $\pi$ pulses driving the
atoms from the excited state down to, and back up from, an auxiliary
third level.  Finally the pulse exits the ensemble in the backward direction as a
time reversed copy of the input.

In this paper, we show that 100\% efficiency using CRIB is theoretically possible using only two level
atoms. Furthermore the output pulse propagates in the same direction as
the input pulse.  It had been thought that such two-level schemes would not be
able to approach unit efficiency because of the problem with re-absorption.  We will show, however, that this is not the case provided that the inhomogeneous broadening is introduced in a way that the detuning of the atoms
is linear with position.  The principle benefit of such a two-level scheme
is in its simplicity.   Firstly, the absence of phase matching $\pi$ pulses greatly
simplifies the implementation.  The only light seen by the atomic ensemble during the operation of the memory is the light field of interest.
Secondly, the memory requires only two atomic levels making this
scheme applicable to many more atomic systems.  In particular erbium dopants
which allow operation at the tele-communication wavelength of 1.5~$\mu$m, have been shown to have very good
two-level characteristics \cite{bott06}, whilst a lambda system has yet to be
demonstrated.
\begin{figure}[!ht]
  \centering
  \includegraphics[width=\columnwidth]{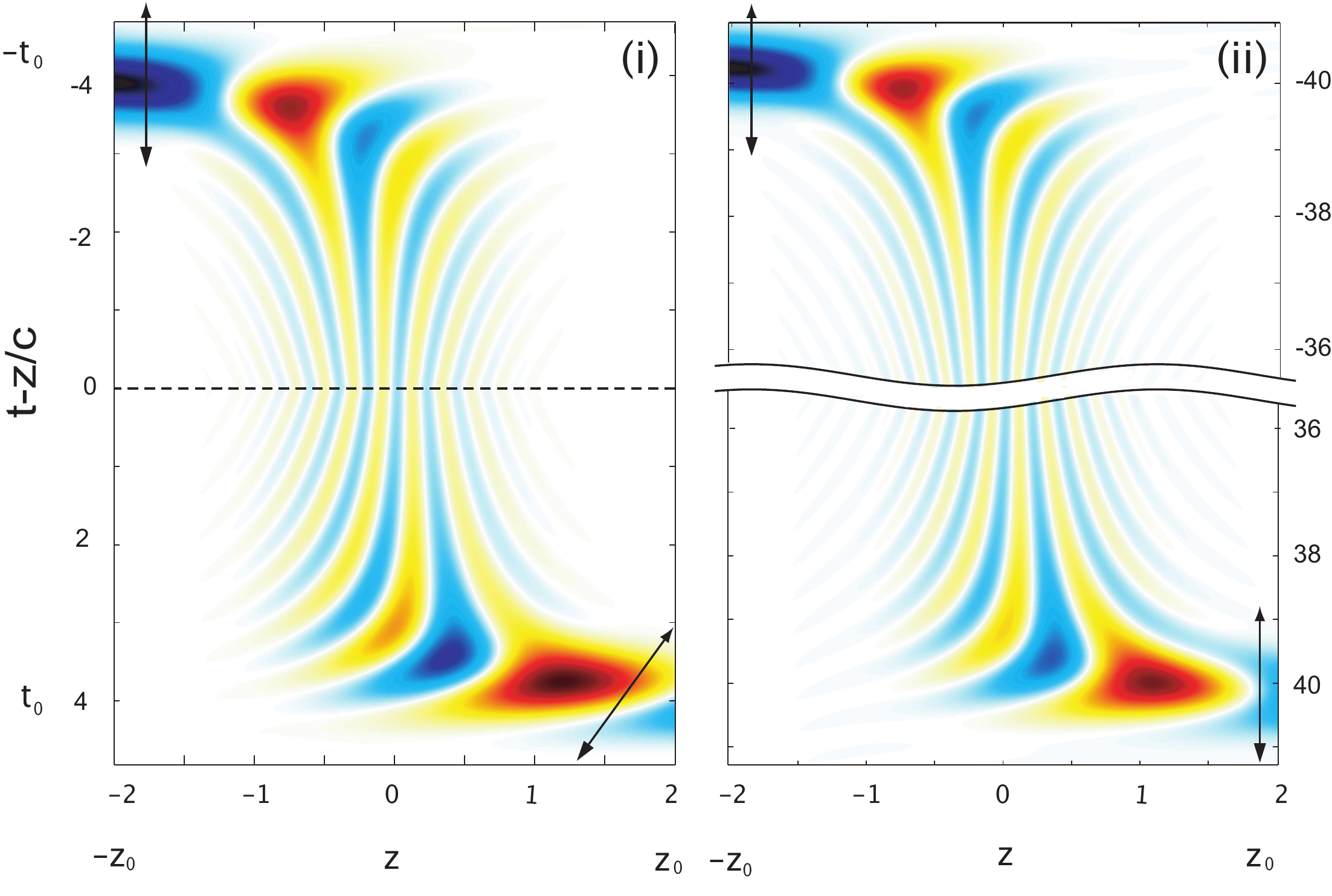}
 \caption{  (Color online) Contour plot of the real part of the optical field in a moving frame at the speed of light. From $t = -t_{0}$ to $t = 0$, the light field enters the sample and is absorbed by the medium.  At $t = 0$, the quadrupole field is flipped and the time reverse process commences producing a forward propagating pulse.  For the same parameters as given in Fig.~\ref{fig:mbsim}, with a storage time 8$\times$ the pulse duration (i) shows a small phase shift across the retrieved pulse.  With a storage time 80$\times$ the pulse duration, (ii) shows a near ideal pulse retrieval. The arrows denote the wavefront of the light fields.}
  \label{fig:realE}
\end{figure}

We consider the interaction of a collection of two-level atoms
with a light field where the detuning of the atoms is linearly dependent on
their position. After the small pulse
approximation, which ensures that a negligible amount of atoms reaches the excited state, the Maxwell-Bloch equations in
the frame at the speed of light are \cite{crisp}
\begin{eqnarray} \label{eq:mb1}
  \frac{\partial}{\partial t} \alpha (z,t) & = & -(\gamma+i \eta z) \alpha(z,t) +igE(z,t)\\
  \label{eq:mb2}
  \frac{\partial}{\partial z}E(z,t) & = &i N \alpha(z,t)
\end{eqnarray}
Where $E$ represents the slowly varying envelope of the optical field; $\alpha$ the
polarisation of the atoms; N the atomic density;  g the atomic transition coupling strength; 
$\gamma$ the decay rate from the excited state and $\eta z$ the detuning from resonance.  
Although the treatment here is classical, the linearity of Eqs.~(\ref{eq:mb1},\ref{eq:mb2}) ensures that the results will hold for quantised fields.  

As shown in Fig.~\ref{fig:mbsim} a), the pulse enters the medium at $z=-z_0$, $t = -t_0$.  The detunings of the atoms is flipped ($-i\eta z \rightarrow +i \eta z$) at $t=0$.  We will show that this leads to a forward propagating output pulse that is closely related to the input at  $z=z_0$, $t = t_0$. Fig.~\ref{fig:mbsim} b) shows the results of numerical simulations for the gradient echo scheme.  For these simulations and the analytical discussion below, we assumed a zero decay rate $\gamma$ and a large optical depth $gN/\eta=10/3$ with an inhomogeneous broadening two times larger than the spectral width of the input pulse.  In practice such a narrow linewidth two-level atom can be realized by optical pumping to a third level that does partake in the quantum memory process.  Under these conditions, the spectral components of the pulse are absorbed at different positions in the sample, which results in a temporal stretching of the pulse as it propagates. When the sign of the inhomogeneous broadening is reversed, the atoms re-radiate the individual spectral components as they re-phase.  Because of the monotonicity of the Stark shift, the time reversed copy of the input pulse propagate in the forward direction with no re-absorption.

Fig.~\ref{fig:realE} shows a contour map of the real part of the electric field.  Because of the large phase shift seen by the field, the last atoms in the sample absorb the
field a long time after the first atoms.  When the switching time is short relative to the pulse duration, a time dependent phase shift is present across the output pulse causing it to be frequency shifted with respect to the input pulse.  Fig. ~\ref{fig:realE}(i) shows results where the storage time is eight times the pulse duration.  For long switching time, nearly all the light field is imparted onto the atomic ensemble.  This leads to a uniform phase shift for the entire pulse and the production of a replica of the input field as shown in Fig. ~\ref{fig:realE}(ii).

To gain a better understanding of the process we present an analytic
treatment which gives an expression for the output field as a function
of the input in the case of large optical depth. First we solve solve
Eqs~(\ref{eq:mb1},\ref{eq:mb2}) subject to the boundary conditions of
$\alpha(z,t\rightarrow-\infty) = 0$ and $E(z=-z_0,t<0) =
f_{\rm{in}}(t)$ to find the values for $\alpha$ and $E$ in the region
($t\in(-\infty,0]$, $z\in[-z_0,z_0]$). To find the values for $\alpha$
and $E$ in the region ($t\in[0,\infty)$, $z\in[-z_0,z_0]$) one needs to
solve a modified version of Eqs~(\ref{eq:mb1},\ref{eq:mb2}) where the
sign of the $i\eta z$ term is reversed. 
One way to do this would be to propagate these equations forward. 
However we will take a
different approach which makes better use of the symmetry of the
protocol. We will guess a particular form of the final
conditions, that is $E(z=z_0,t>0) = f_{\rm{out}}(t)$,
$\alpha(z,t\rightarrow\infty)=0$ and propagate these expressions 
backwards to find $\alpha(z,t=0)$, $E(z,t=0)$.  We will then validate that 
our guess is correct by showing that the boundary conditions for our backward
propagated solutions for the region $t>0$ match those that we found
for the region $t<0$. Integrating Eq.~(\ref{eq:mb1})
gives
\begin{equation}
 \alpha(z,t)  = i g \int_{-\infty}^\infty dt' \, H(t-t')e^{-i\eta z(t-t')}
 E(z,t') 
\end{equation}
where $H(\xi)$ is the Heaviside step function. Fourier transforming
this expression, substituting into Eq.(\ref{eq:mb2}) and then integrating
the result gives  
\begin{eqnarray} \label{eq:soln1}
  \tilde{E}(z,\omega) &=& \tilde{F}_{\rm{in}}(\omega) \exp[ -\pi\beta(H(\omega+\eta z)-H(\omega-\eta z_0)) \nonumber \\
  &&  i\beta\ln\left |\frac{\omega+\eta z}{\omega-\eta z_0}\right| ] 
\end{eqnarray}
Here $\tilde{F}_{in}(\omega)$ is the Fourier transform of $f_{in}(t)$ and
$\beta=gN/\eta$ is the optical depth.
It can be seen that each spectral component is
attenuated by a factor $\exp(-\beta)$ after travelling past the
position in the sample where it is resonant with the atoms, as well as
getting a phase shift as it travels through the sample.

Because of this phase shift, the intensity of the field tends to zero with time asymptotically.
Its expression at $t=0$, can be obtained by integrating Eq.(\ref{eq:soln1}) 
over all $\omega$, where we assume that the
bandwidth of the memory is larger than the bandwidth of the input
pulse,  $\eta z_{0} \gg \omega$. This simplification leads to an
expression for $E(z,t=0)$ in the form of a convolution. Fourier
transforming with respect to the spatial coordinate leads to 
\begin{eqnarray}
E(k,t=0) & = &-f_{in}(-\frac{k}{\eta})\sgn(k)\beta  \left|\frac{k}{\eta}\right|^{-2-i \beta } \Gamma (i \beta ) \nonumber \\
             &&  \left(\left|\frac{k}{\eta}\right| \cosh \left(\frac{\pi
    \beta }{2}\right)+\frac{k}{\eta} \sinh \left(\frac{\pi  \beta }{2}\right)\right)   \label{eq:kspace}
\end{eqnarray}
Here $f_{\rm{in}}(k/\eta)$ is the
input field at the time $\tau=k/\eta$ and $\Gamma(\xi)$ is the gamma function. 
 
As mentioned above we can also derive an expression for $E(k,t=0)$ subject
to the final conditions of $\alpha(z,t\rightarrow\infty) = 0$ and $E(z=z_0,t>0)
= f_{\rm{out}}(t)$, one gets a similar expression to Eq.(\ref{eq:kspace}) but in terms of $f_{\rm{out}}(k/\eta)$. The backwardly propagated
solution satisfies the boundary condition $E(z=-z_0,t>0)=0$ and $\alpha(z,t\rightarrow\infty) = 0$ so long as
the optical depth is sufficiently high. 
Using continuity arguments, we match the two solutions found for $E(k,t=0)$ and get an expression of the output field at a time $t_0$.

\begin{equation}
  \label{eq:tada}
  f_{\rm{out}}(t_0) = f_{\rm{in}}(-t_0) |t_0|^{2i\beta}\frac{\Gamma(i \beta)}{\Gamma(-i \beta) }.
\end{equation}

Because $|t_0|^{2i\beta}{\Gamma(i \beta)}/{\Gamma(-i \beta) }$ has a
modulus of 1 for all $t_0$, it can bee seen that the envelope of the
output pulse is a time reversed version of the input. 
When the length of the pulse is small compared
to the total delay it is also phase shifted by $|t_0|^{2i\beta}$. This term describes a frequency shift of 
the output pulse with respect to the input.
One other elegant way to cancel this frequency shift
 is to cascade two gradient echo memories where the flipping is performed the opposite way for the second memory.
It can be see from Eq.(\ref{eq:tada}) that the frequency shift from each memory would indeed cancel. 

\begin{figure}
  \centering
  \includegraphics[width=\columnwidth]{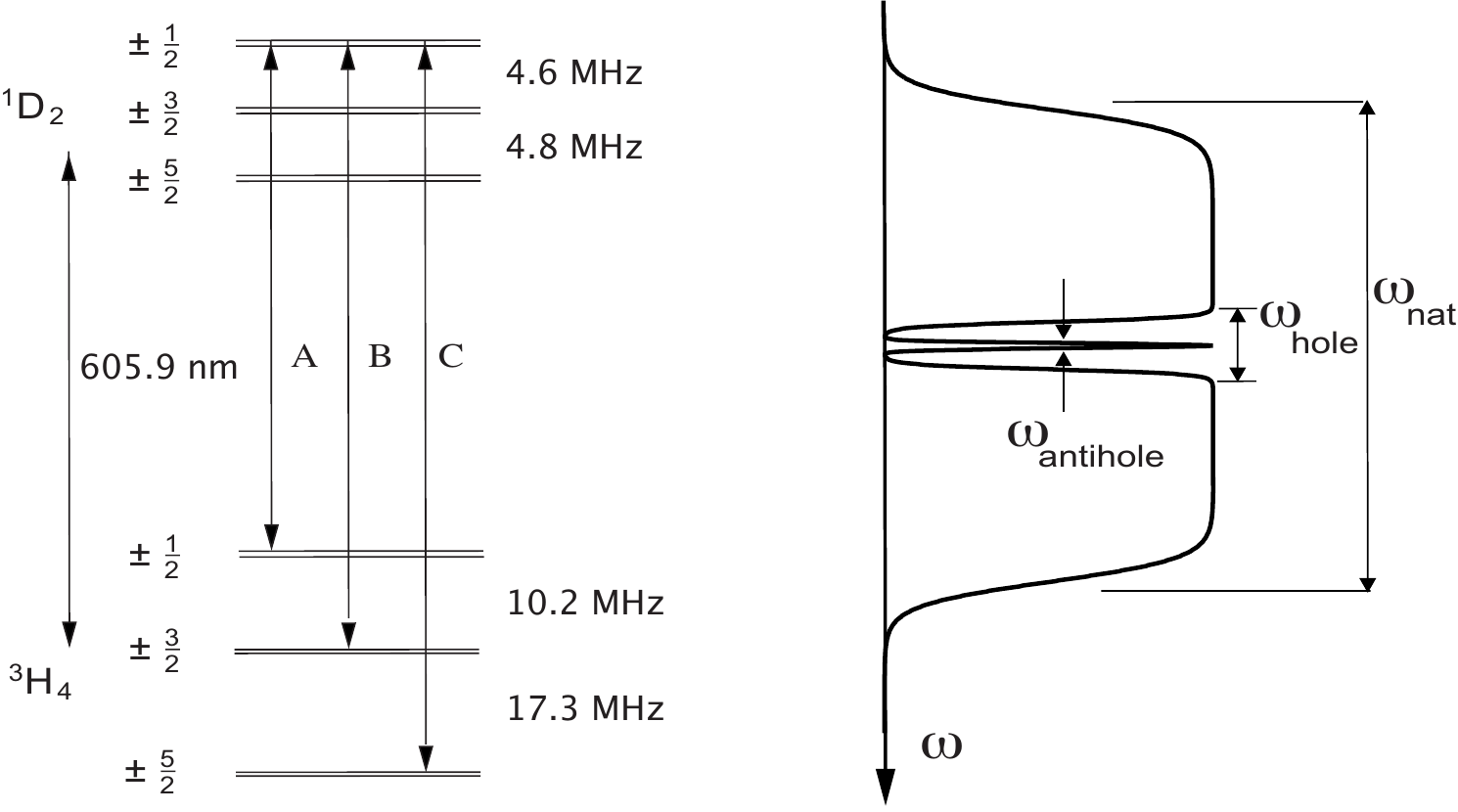}
  \caption{Energy level diagram and spectral scheme for the praseodymium dopants in yttrium orthosilicate.  The gradient echo is carried out with light at frequency A.  The natural inhomogeneous linewidth of the sample, $\omega_{\rm nat}$, is a few GHz wide.  To set up the experiment, the applied light is swept around frequency A to create a spectral hole a few MHz wide, $\omega_{\rm hole}$.  Light at frequency B and C is then applied to prepare a narrow antihole around A with linewidth $\omega_{\rm anti}$.  Although the diagram shows the use of the $\pm 1/2$ excited states, there are ions contributing to the antihole from the $\pm 3/2$ and $\pm 5/2$ excited states due to inhomogeneous broadening in the optical transition.  In our experiment $\omega_{\rm antihole} =$ 30 kHz.}
  \label{fig:level_diag}
\end{figure}

\begin{figure}[ht!]
  \includegraphics[width=\columnwidth]{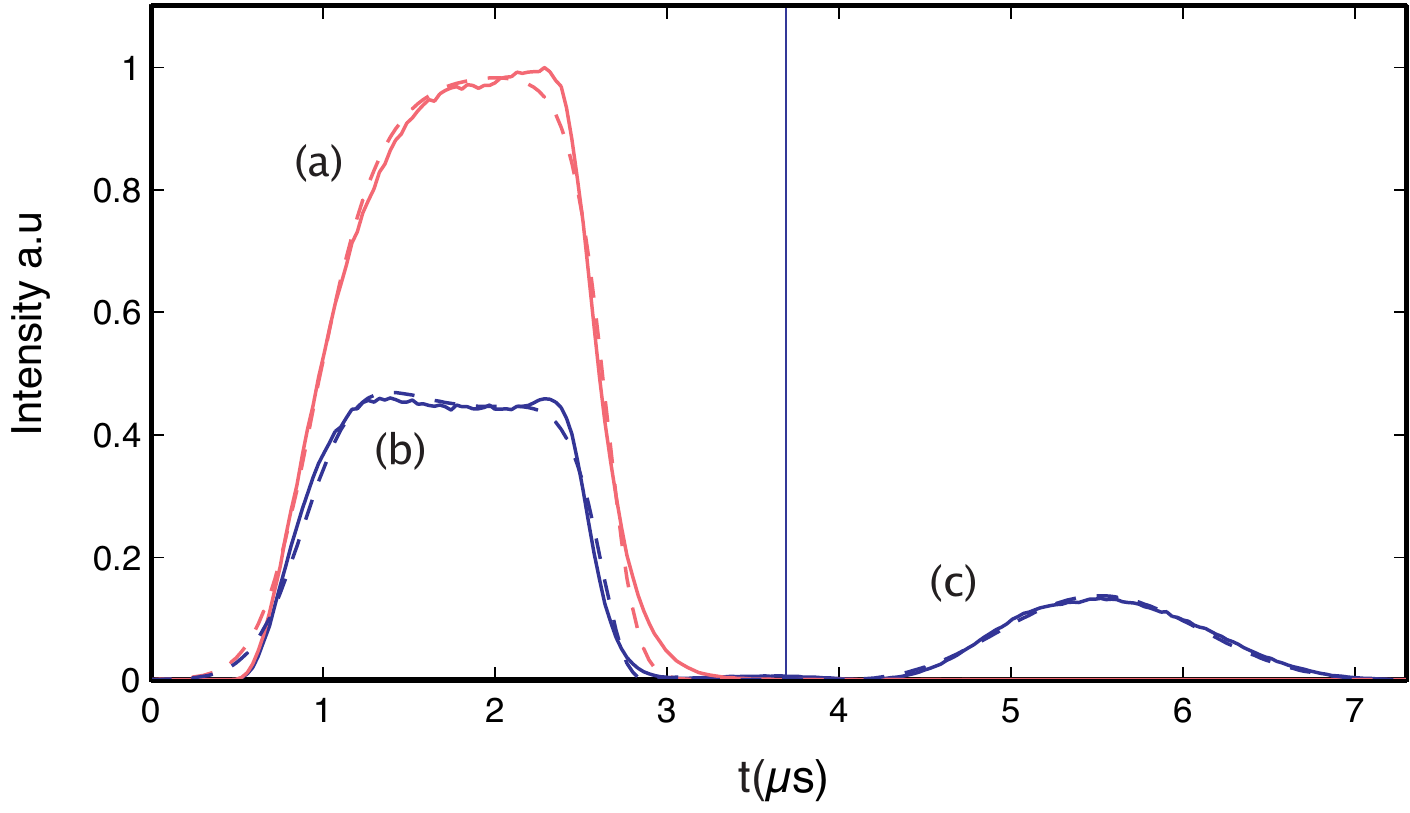}
  \centering
  \caption{Experimental results (solid lines) and the numerical simulations (dash lines) of the gradient echo setup.  When no antihole is prepared, trace (a) shows the input pulse being detected after transmission through the sample.  The gradient echo trace shows (b) the transmitted and (c) the stored-and-recalled pulse components.  The vertical line at 3.7 $\mu$s denotes the time at which the Stark shifting electric field is flipped.}
  \label{fig:echo_trace}
\end{figure}

The theoretical treatment presented above showed that this Gradient Echo scheme can in principle 
be used as a quantum memory for light.
Preliminary experiments have also been carried out in a similar manner to
the initial demonstrations of photon echoes produced using controlled
inhomogeneous broadening \cite{alex06}. The most significant
difference was the use of praseodymium rather than europium dopants,
which provided much greater optical depths. The
experiment was carried out on a spectral antihole which was prepared
as described in Fig.~\ref{fig:level_diag}.  Light from a highly
stabilised dye laser was frequency shifted and gated with
acousto-optic modulators (AOMs). The pulse was then steered toward the sample of
0.05~atomic~\% praseodymium doped in yttrium orthosilicate. The sample was
approximately a 4~mm cube and was held at temperatures in the range
2-4~K. Four electrodes were placed around the sample in a quadrupole
arrangement and provided an electric field that varied linearly along
the optical path. The electrodes were 1.7~mm diameter rods separated
by 8~mm. Voltages of approximately $\pm$5~V were used to broaden the
antihole. Heterodyne detection was used to detect the transmitted
pulses.

Fig.~\ref{fig:echo_trace} shows the typical experimental traces of the gradient echo memory with and without the preparation of the two-level anti-hole.  In the results, 49\% of the incident light was transmitted straight through the sample and 13\% of light was recalled as
an echo.  Due to some variation in the antiholes created, we observed a variation of the efficiency of the echo between the values of 10 and 15\%.  Using our numerical model, we vary the spectral width of the unbroadened anti-hole and the optical depth of the sample to match the experimental results.  Close agreement between the experimental results and the simulations is obtained only with these two free parameters.  Our numerical model suggests an anti-hole width of 30~kHz.  This is again in agreement with the experimental expectation, if the hyperfine transition broadening were the main limitation to the antihole linewidth.
For a given pulse length, the optimization of the experiment is dependent on a compromise between maximizing the optical depth of the sample and increasing the linewidth ratio between the applied and the intrinsic inhomogenous broadenings. In our experiment, $gN/\eta = 0.006$, whilst the ratio between applied and intrinsic linewidth is around 200.

For each crystallographic site where praseodymium is located, there is
another related to it by inversion. In order to implement a completely efficient
memory only one of the site pair can be used.  In principle, this could be achieved by Stark shifting
with a homogeneous electric field and optical pumping.  In our experiment, however, both orientations were used.  The theoretical modeling on Fig.~\ref{fig:echo_trace} takes into account these two orientations by having two Bloch equations and two source terms for the optical field. 
The performance of the gradient echo memory can be enhanced with minor improvement to the set-up.
Simulations suggest that with Fourier limited pulses, selecting only one 
orientation of the praseodymium ions and increasing the optical depth by a factor of 
three (for instance, by increasing the crystal length) would enable the scheme to reach more than 50\% efficiency.

In conclusion we have proposed a quantum memory for light based on
optical gradient echoes.  When compared to existing quantum memories
based on controlled inhomogeneous broadening, our scheme requires
only two atomic levels and is therefore easier to implement on a wider range of systems.  Moreover, our scheme does not require optical pulses for the storage and recall process.  Initial experiments show an average echo efficiency of 13\% and a time bandwidth product of around three.  This compares favourably with the performance of quantum memories based on EIT
\cite{chan05,eisa05}. The experiments are well modeled by simple Maxwell-Bloch equations.  Simulations of modest improvements on the experimental parameters suggest that efficiency higher than 50\% should be realizable.


\end{document}